\documentclass[nonacm,sigplan]{acmart}

\AtBeginDocument{%
  }

\setcopyright{acmlicensed}
\copyrightyear{2024}
\acmYear{2024}
\acmDOI{XXXXXXX.XXXXXXX}
\acmConference[ASE'24]{39th IEEE/ACM International Conference on Automated Software Engineering}{Oct 27--Nov 1,
  2024}{Sacramento, CA}
\acmISBN{978-1-4503-XXXX-X/18/06}

\begin{document}

\title{Test-Driven Development for Code Generation}

\author{Noble Saji Mathews}
\orcid{0000-0003-2266-8848}
\affiliation{%
  \institution{University of Waterloo, Canada}
   \country{}
}
\email{noblesaji.mathews@uwaterloo.ca}

 \author{Meiyappan Nagappan}
\affiliation{%
  \institution{University of Waterloo, Canada}
   \country{}
}
\email{mei.nagappan@uwaterloo.ca}

\begin{abstract}
Recent Large Language Models (LLMs) have demonstrated significant capabilities in generating code snippets directly from problem statements. This increasingly automated process mirrors traditional human-led software development, where code is often written in response to a requirement. Historically, Test-Driven Development (TDD) has proven its merit, requiring developers to write tests before the functional code, ensuring alignment with the initial problem statements. Applying TDD principles to LLM-based code generation offers one distinct benefit: it enables developers to verify the correctness of generated code against predefined tests. This paper investigates if and how TDD can be incorporated into AI-assisted code-generation processes. We experimentally evaluate our hypothesis that providing LLMs like \textit{GPT-4} and \textit{Llama 3} with tests in addition to the problem statements enhances code generation outcomes. We experimented with established function-level code generation benchmarks such as MBPP and HumanEval. Our results consistently demonstrate that including test cases leads to higher success in solving programming challenges. We assert that TDD is a promising paradigm for helping ensure that the code generated by LLMs effectively captures the requirements.
\end{abstract}

\ccsdesc[500]{Software and its engineering~Software development techniques}
\ccsdesc[300]{Computing methodologies~Artificial intelligence}

\keywords{Code Generation, LLM, TDD, Testing, Software Engineering}

\newcommand\RqOne{RQ1}
\newcommand\RqTwo{RQ2}
\newcommand\RqThree{RQ3}
\newcommand\RqFour{RQ4}
\newcommand\RqFive{EXP1}
\newcommand\RqSix{EXP2}
\newcommand\RqSeven{EXP3}

\definecolor{arsenic}{rgb}{0.23, 0.27, 0.29}
\definecolor{azure}{rgb}{0.0, 0.5, 1.0}
\newcommand\RqOneText{How good is basic code generation? }
\newcommand\RqTwoText{Does providing test cases improve code generation?}
\newcommand\RqThreeText{Can failed tests help LLMs fix their mistakes?}
\newcommand\RqFourText{Do the generated solutions only satisfy the supplied tests or even unseen tests?}

\newcommand\RqFiveText{How does problem difficulty affect these results?}
\newcommand\RqSixText{How many tests is good enough?}
\newcommand\RqSevenText{Do these results hold with an open model like Llama 3?}

\maketitle

\section{INTRODUCTION}
Large Language Models (LLMs) are transforming software development. However, the importance of correctness in this domain remains a challenge. Despite their advanced capabilities, LLMs often exude deceptive confidence in their outputs, which can be misleadingly erroneous \cite{spiess2024quality, liu2024your, virk2024enhancing}. This issue becomes especially critical when the cost of failure is high. The potential pitfalls of LLMs, such as generating syntactically correct but logically flawed code or failing to adhere to specific requirements, underscore the need for stringent validation mechanisms. Ensuring correctness is not just about preventing errors. It is about building trust in machine-generated code and ensuring it meets the rigorous standards required in real-world applications. Programmers will need to act as quality assurance specialists, spending more time checking the validity of autogenerated code, utilizing both traditional testing tools and advanced automated program repair techniques to ensure code quality \cite{lyu2024automatic}

Generating tests has gained additional momentum with the emergence of LLMs, and recent techniques boast testing performance similar to human-written tests. Past work, however, has questioned how coverage as a goal may not be an effective target to set \cite{fraser2015does}. If LLMs fail to handle edge cases regarding code, generated tests derived from the implementation can be as wrong as the generated code itself. While LLMs might be useful in helping developers formalize intent, this paper does not seek to answer the question "When and how should AI write tests?". This work rather seeks to explore whether there might be a better paradigm for utilizing tests in LLM-based code generation.

Test-driven development (TDD) \cite{beck2022test} offers a structured approach to mitigate the risks associated with code generation. This software development process revolves around the repetitive cycle of writing a test for a specific function or feature, developing code to pass that test, and subsequently refactoring the code to meet the desired standards of quality and efficiency. Hence, TDD could serve as a critical framework for validating the correctness and functionality of LLM-generated code. 

There is extensive research into generating test cases from code, whether human or machine-authored \cite{schafer2023empirical, lemieux2023codamosa}. However, the reciprocal process has not been systematically evaluated, particularly regarding how human-written tests (created as part of the TDD process) can improve LLM-generated code. By examining how LLMs respond to the additional context and constraints of test cases, we can gain insights into their potential to enhance code correctness through a more informed code generation process. 
While it may seem obvious to assume that including tests alongside problem statements will inherently enhance the correctness of machine-generated code, this assumption requires careful examination. The actual impact of human-written tests on improving the quality of LLM-generated code remains underexplored. Specifically, the extent to which LLMs can interpret and utilize these tests to refine their code outputs has not been thoroughly investigated. This gap highlights a critical area for empirical study: measuring the effectiveness of test cases in guiding LLMs toward generating more accurate code that captures requirements which may not be explicitly stated~(much like the goals of TDD).

In our empirical study, we create a simple framework \textit{TGen} to evaluate our research questions. \textit{TGen} is rooted in TDD principles, requiring developers to precisely articulate their coding intents through tests. \textit{TGen} operates within a continuous self-improving loop. Developers begin by detailing their requirements and intended functionalities through a combination of problem statements and test cases. These test cases then serve as a definitive guide for the expected behaviour of the code, allowing \textit{TGen} to generate code that adheres to these requirements. Integral to the framework is an LLM-based feedback mechanism designed to iteratively refine and correct the generated code, ensuring it passes all provided unit tests. By leveraging test cases as a benchmark for validation, \textit{TGen} attempts to generate code satisfying these requirements. In cases where \textit{TGen} cannot create correct code, by design, developers know that the code generated does not pass all the test cases and, therefore, can correct the code before including it in their projects.


We investigate whether incorporating tests in the code generation process improves the model's ability to generate correct code and also discuss the impacts and challenges of this approach. While we use \textit{TGen} in our study for evaluation, it’s important to note that our contribution is not the framework itself but the empirical study we conduct. We systematically explore whether the TDD paradigm works in AI-driven code generation, under what conditions it is effective, and how tests can be useful. Our findings can support or challenge the use of TDD in this context.

By integrating TDD, developers can continuously test the outputs of LLMs against predefined specifications and use cases. This helps identify and correct errors early in the development process and aligns the model’s outputs with the specific requirements of the task at hand. Thus, the adoption of TDD for code generation could stand as a promising strategy to enhance the practical utility of these advanced AI systems in software development.

\section{Research Questions}

Our goal is to examine the baseline capabilities of neural code generation and the incremental benefits introduced by incorporating tests. Specifically, we explore if and how tests are helpful through the following four research questions:

\begin{enumerate}

\item [\textbf{\RqOne{}:}] \textbf{\RqOneText{}}
Basic neural code generation, particularly with LLMs, has shown promising results in various tasks. However, its efficacy in complex coding scenarios without any test information remains a subject of exploration. We aim to evaluate the fundamental capabilities of these models in generating syntactically and logically correct code, especially in standard programming tasks. This investigation will provide insights into the baseline performance of LLMs in code generation, setting a benchmark for further discussions.

\item [\textbf{\RqTwo{}:}] \textbf{\RqTwoText{}}
The foundational principle of TDD is that software should be developed primarily by writing tests before any functional code is written. This method has historically led to robust, well-documented software that is closely aligned with user requirements. By requiring developers to define explicit tests upfront, TDD ensures that all new code functions as intended and adheres strictly to the specified requirements. This rigorous approach to software development could be transformative when applied to the code-generation capabilities of LLMs. We hypothesize that by providing LLMs like GPT-4 with both problem statements and corresponding tests, the generated code will achieve the required functionality and exhibit an enhanced adherence to the intended specifications and requirements.

\item [\textbf{\RqThree{}:}] \textbf{\RqThreeText{}}
We investigate the ability of LLMs to learn from their errors through remediation loops. This involves assessing whether LLMs can iteratively refine and correct their code outputs based on feedback or error detection. The remediation loop in our study closely mirrors the iterative cycle at the heart of TDD. In traditional TDD, the process involves initially writing tests, which the new code then fails, leading developers to refine the function until it passes the tests. We aim to explore the potential of LLMs to self-correct as reported in past literature \cite{weng2022large}and the ability to align better the code they generate with the provided specifications.

\item [\textbf{\RqFour{}:}]\textbf{\RqFourText{}}
This research question explores a critical aspect of AI-generated code: its robustness and generalizability beyond the explicitly defined test cases. In traditional software development, the ability of code to perform well under a variety of scenarios—especially those not explicitly anticipated during testing—is a hallmark of high-quality software. TDD is grounded in the principle that tests are pivotal in improving design and functionality. However, there remain questions about the extent to which this principle applies when code is generated by LLMs.

\end{enumerate}

In addition to these research questions, we delve deeper into the complexities of incorporating tests, discussing our findings further through the following additional explorations:

\begin{enumerate}

\item [\textbf{\RqFive{}:}]\textbf{\RqFiveText{}}
We hypothesize that a problem statement's difficulty could significantly impact an LLM's performance in code generation. We aim to explore how varying levels of problem complexity influence the model's ability to generate correct code that satisfies the requirements expressed by the statement and supplied test cases. We curate our own dataset consisting of competitive programming problems of different difficulty levels and attempt to understand how increasing difficulty impacts our approach. 

\item [\textbf{\RqSix{}:}]\textbf{\RqSixText{}}
Determining the optimal number of tests necessary for effective code generation by LLMs is a key concern. In traditional TDD, the number and quality of tests are crucial for ensuring robust software. We aim to investigate how the quantity of provided test cases influences the correctness of the code generated by LLMs. While intuitively, adding tests should only help clarify the requirements. However, when working with LLMs, we are limited by the context length of the model, which controls how much information we can feed the model. By experimenting with varying numbers of tests, we wish to identify if there exists a threshold that balances thoroughness and efficiency, guiding best practices for using tests to enhance LLM-based code generation.

\item [\textbf{\RqSeven{}:}]\textbf{\RqSevenText{}}
While our approach is LLM agnostic, the LLM itself plays a major role in our experiments, requiring the ability to write correct code and review and fix incorrect code. While there may exist an ideal combination of different LLMs specialized for each of these tasks, we wish to explore if this structure would work with other models outside OpenAI's GPT family of models as well. By evaluating Meta's LLama 3 under the same conditions, we can assess if similar patterns emerge, providing insights into the consistency and reliability of our findings across different LLMs. 

\end{enumerate}

\section{METHODOLOGY} 

\begin{figure*}[ht]
    \centering
    \includegraphics[width=\linewidth]{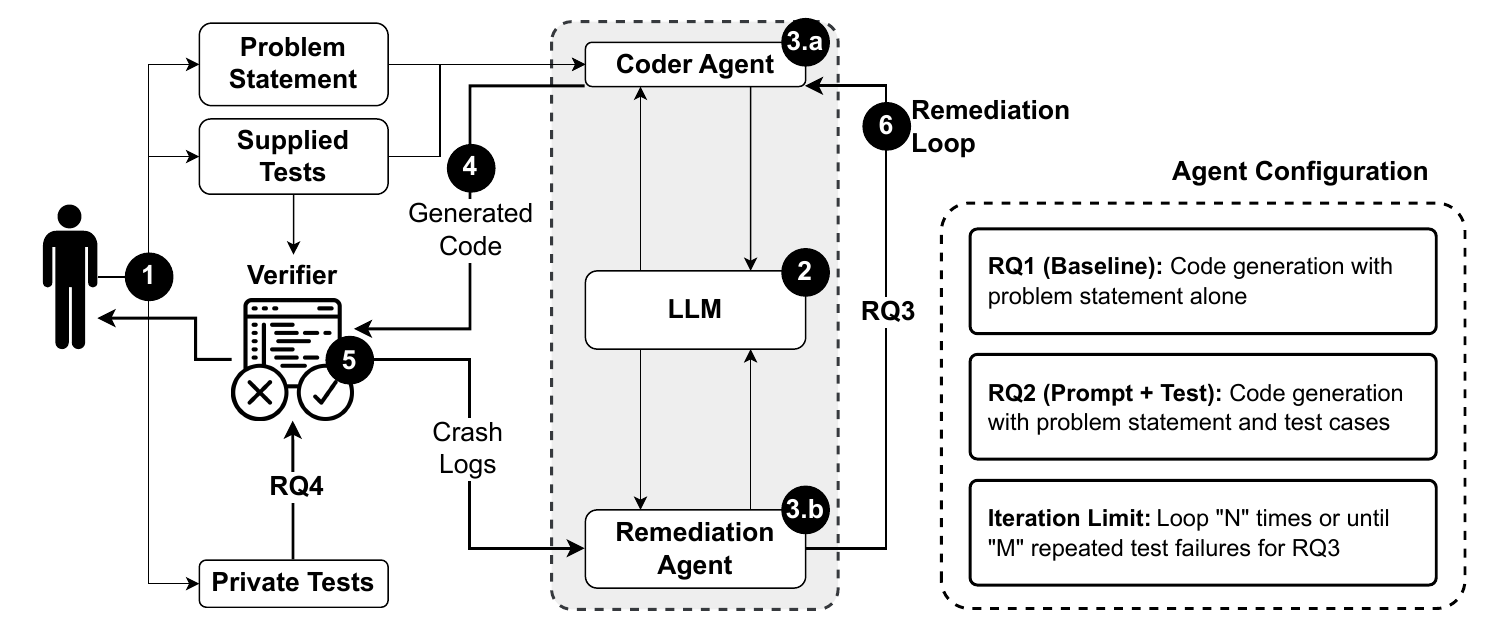}    
    \caption{Overview of the evaluation pipeline}
    \label{fig:Approach}
\end{figure*}

We design our experiments to help evaluate the effectiveness of integrating TDD with LLM-based code generation. We aim to determine if tests affect LLMs and when they might be useful.

\subsection{Datasets Used}

Evaluating code generation typically employs benchmark datasets such as MBPP \cite{austin2021program}, HumanEval \cite{chen2021evaluating}, CodeContests \cite{li2022competition}, or APPS \cite{hendrycksapps2021}. Recently, newer datasets have explored class-level \cite{du2024evaluating} and repository-level code completion \cite{zhang2023repocoder}, as well as automated software engineering \cite{jimenez2023swe}. However, we focus on function-level generation datasets like MBPP and HumanEval for two reasons. 
\begin{enumerate}
    \item MBPP and HumanEval datasets are standard benchmarks widely used to evaluate LLMs on code generation leaderboards \cite{du2024evaluating, liu2024your}.
    \item Repository-level code generation requires context about the codebase, making it complex and less controlled. Class-level benchmarks can be evaluated in various generation configurations, introducing many variables complicating the analysis. Function-level generation datasets like MBPP and HumanEval are relatively more straightforward. These involve generating the body of a function with a given signature, providing a clear and manageable scope for evaluation. By focusing on MBPP and HumanEval, we can systematically explore our research questions in a controlled environment where we can focus on TDD and its effects.
\end{enumerate}

 By establishing the effectiveness of TDD in LLM-based code generation using these datasets, we lay the groundwork for exploring more complex datasets focused on real-world TDD data in future studies. Conversely, if TDD proves ineffective with these benchmarks, it may indicate limited applicability in this context. We utilized refined versions of HumanEval and MBPP datasets curated by EvalPlus \cite{liu2024your}, which adds additional tests to allow for a more rigorous evaluation of the generated code.

\begin{itemize}
    \item \textbf{HumanEval}: Contains 164 handwritten Python programming problems. Each problem includes a function signature, docstring, body, and several unit tests. The EvalPlus variant we use includes additional unit tests (80 times the original) for each of the 164 problems, providing more extensive coverage and thorough testing.
    
    \item \textbf{MBPP}: Comprises around 1,000 crowd-sourced Python programming problems designed for entry-level programmers. Each problem includes a task description, a code solution, and three automated test cases. The version we use includes 35 times the original number of unit tests for each problem. It consists of 399 tasks, which are a subset of the original MBPP dataset. This subset is selected by the authors of EvalPlus from the sanitized MBPP (a set of 427 manually filtered by the original MBPP authors), with further removal of low-quality and ill-formed tasks for benchmark quality control.
\end{itemize}

We never supply the LLM with any additional test cases that EvalPlus added using dynamic program testing techniques and only use them to test correctness. However, existing work in the literature has questioned the accuracy of function-level benchmarks for evaluating LLMs \cite{roziere2023code, cassano2023multipl}. We also look at file-level generation to address this concern and validate our findings. For this, we curate our own dataset of 1100 problems from a popular competitive programming website, CodeChef, for the following reasons: 
\begin{itemize}
    \item We also wish to explore how difficulty impacts our results, and existing datasets do not have a balanced set of problems spanning various difficulty levels
    \item CodeContests v2 is not public, and the older dataset is outdated and filled primarily with Python 2 solutions and problems that no longer accept solutions on the competitive programming platforms they were extracted from
    \item APPS dataset, which is a more recent dataset of coding challenge problems, does not give a clear difficulty measure but rather categorizes problems into abstract buckets, which doesn't work well for our purposes
\end{itemize}

\subsection{TGen Framework}

\begin{figure}[t]
\centering
\begin{minipage}{\linewidth}

\hrulefill

Your task is only to write a Python function to satisfy requirements specified by the users Prompt.

Only generate a single complete Python code snippet. Start generating after the [PYTHON] tag. Your python solution should end with a [/PYTHON] tag.
\\

{
\textit{
$|$ Multi-step prompting for case with tests supplied}
\\

1. Look at the "Prompt" and "Tests" provided to understand the users requirements.

2. Generate code based on requirements in the prompt and ensure the logic is such that it would pass the corresponding tests provided.

3. Do not attempt to programmatically verify the tests. The test should not be part of the generated code in any way.

4. Do not write code in order to satisfy the tests, primarily focus on satisfying the prompt.

5. Include necessary imports in the generated code.

6. Your code is expected to read input from stdin and write output to stdout as per the requirements in the prompt.

7. Define additional helper functions if needed

8. Your output should be as shown in the sample tests. Ensure you do not return or print any additional information that can cause the tests to fail.
}
\\

{
\textit{$|$ Dataset-specific guidelines for function-level case
}
\\

If starting code is provided look at the issues mentioned and attempt to fix them

Ensure that the signature of the function is the same as the one provided by the user

And format your response as follows:

[PYTHON]

def function\_name(parameters\_as\_provided\_in\_signature):

\{

Ensure the function signature is the same as the one provided in the signature

Infer meaning of parameters from the prompt and tests

Valid python code with necessary imports if any that satisfies the prompt

Ensure the entry method is a function with the same name as the function name in the tests

\}

[/PYTHON]
}

\hrulefill

\caption{Prompt used by the Coder Agent}
\label{fig:generation_prompt}
\end{minipage}
\end{figure}

\begin{figure}[t]
\centering
\begin{minipage}{\linewidth}

\hrulefill

\textbf{System:} You are an expert Python programmer, your job is to look at code and reason why it doesn't work as intended. Once you reason why it doesn't work, you will provide a prompt that highlights the key points that need to be fixed in the code. You will not write the code yourself. IMPORTANT: Tests cannot be modified. Never make a suggestion outside the scope of modifying the solution code snippet

\textbf{Assistant:} 
\textit{
<Code Snippet under remediation>
}

\textbf{User:} This code is not correct as it led to the following issues:

\textit{
<List of issues following test execution extracted by log-analyzer>
}

\hrulefill

\caption{Configuration of the Remediation Agent}
\label{fig:remediation_prompt}
\end{minipage}
\end{figure}

Figure \ref{fig:Approach} provides an overview of the proposed \textit{TGen} framework’s pipeline, which we use for evaluation. Each part of the pipeline is described below:
\begin{enumerate}
    \item \textbf{Input Phase}: The evaluation pipeline starts by accepting two main inputs: the \textit{problem statement} describing the task and the \textit{tests} containing the public unit tests and the required signature, if any, to be enforced in the desired solution. These tests serve both as specifications for the code to meet and as a verification tool for the output. These tests serve as both specifications and verification tools. Tests provided to the LLM during generation are labelled as supplied tests, while the rest are private tests. For competitive programming problems, the challenge description is the problem statement, and available public tests are supplied tests.
    
    \item \textbf{LLM Engine}: The core reasoning engine, powered by state-of-the-art LLMs (e.g., Meta Llama 3, GPT-3.5 \& GPT-4 Turbo), processes the structured input. The foundational model leverages its own knowledge and any supplied context to generate an output response. This study uses GPT-4 Turbo v1106 for all experiments, with results validated using Meta Llama 3.

    \item \textbf{LLM Agents}: In the context of Large Language Models (LLMs), "agents" are systems that utilize LLMs as their underlying reasoning engine \cite{braga2023decentralised}. These agents are designed with predefined prompts and thought patterns, enabling them to interact intelligently with users or other systems. Agents in our pipeline utilize the intent and available context from the input phase to guide the LLM engine in generating code, reviewing snippets, and suggesting fixes.
    \begin{enumerate}
        \item \textbf{Coder Agent}: This agent takes the supplied information and generates a code snippet from it. The system prompt used when tests are supplied for function-level generation cases is shown in Figure~\ref{fig:generation_prompt}. The first part provides a rough thought process to instruct the LLM on how to utilize the tests, and the second part contains instructions specific to the structure of tasks in a dataset. Though the overall structure of the prompt remains the same, variants of this are used in different cases, namely, when no tests are supplied or when a file-level generation task is performed.
        \item \textbf{Remediation Agent}: This agent looks at failures reported by the verifier and attempts to suggest how to fix the code, which is then taken up by the ``Coder Agent'' in the next iteration. The structure of the prompts is shown in Figure~\ref{fig:remediation_prompt}. 
    \end{enumerate}
    
    \item \textbf{Output Phase}: The \textit{Generated Code} is extracted from the LLMs output, which is then subject to validation. The output is expected to satisfy both the functional requirements and the constraints imposed by the tests, if any.
    
    \item \textbf{Verification}: The output of the coder agent is always run through the verifier that uses PyTest, a Python unit-testing framework, to verify the generated code against all available tests. The verifier collects crashes or test failures and creates a comprehensive list of failures for the LLM to inspect and address issues. The verifier identifies any discrepancies and ensures only correct solutions are accepted and returned to the user.
    
    \item \textbf{Remediation Loop}: The system enters a remediation loop if the generated code fails the tests. The remediation agent uses failure information (e.g., stack trace, detected issues, executed lines) to refine feedback for the coder agent. This loop runs for a fixed number of iterations or until the same tests fail repeatedly, showing no progress. We limit the loop to five iterations and three repeated failures for our experiments. We observed that the remediation advice often becomes repetitive beyond three to four iterations, yielding diminishing returns on further attempts to fix the code.
\end{enumerate}

To conduct our experiments in a cost and compute-efficient manner, our pipeline executes in the following fashion:
\begin{itemize}
    \item \textbf{Solved without tests:} The first solution is attempted with the problem statement alone and verified for correctness
    \item \textbf{Needs tests:} If the problem statement alone fails to generate a valid solution, the pipeline is rerun, supplying all public tests along with the problem statement. The newly generated code is then verified against all tests
    \item \textbf{Needs remediation:} If the previous step results in an invalid solution, the failure data is collected from the verifier and the remediation loop is executed until a correct solution is generated
    \item \textbf{Unsolved:} If the remediation loop exits without generating a valid solution, the problem is considered unsolvable by \textit{TGen}
\end{itemize}

We provide a replication package \footnote{\url{https://osf.io/e3jy6/?view_only=bc67e33bebd3435abf5537d56767401d}}, which includes output and runtime details from \textit{TGen} for all cases we experimented with and code scripts for the experiments. Output from LLMs is often unpredictable. We use a seed value of "1106" and a temperature of 0 in our experiments to improve the replicability of our results.

\begin{figure}
    \centering
    \includegraphics[width=1\linewidth]{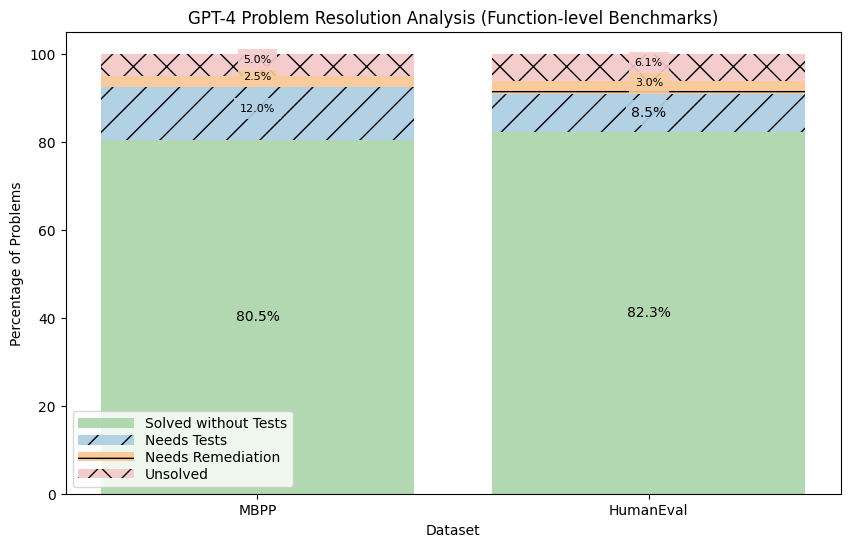}
    \caption{GPT4: Problems resolved and strategies used for Function-level code generation benchmarks}
    \label{fig:gpt4wofunction}
\end{figure}

\section{IMPACT OF TDD IN CODE GENERATION} \label{impact}

In this section, we seek to answer the “IF” question: Do tests help the code generation process? We also manually explore the data to assess the impact of tests and determine which types of problems benefit from tests and which do not. The fraction of correct solutions obtained following the aforementioned workflow with GPT-4 Turbo version 1106 as the LLM for the two function-level benchmark datasets MBPP and HumanEval can be seen in Figure~\ref{fig:gpt4wofunction}. RQs 1 through 4 break down these results further. Satisfying public tests is considered the correctness criterion in RQs 1, 2 and 3. RQ4 dives into the notion of private tests and uncaptured requirements. We also manually investigate the data to highlight key details and insights. Please note that we refer to specific cases within the dataset using the following notation ``Dataset/Key'' where the key is a unique identifier for the problem within the corresponding dataset. 

\subsection{\RqOneText{}}

GPT-4, being one of the most powerful LLMs available today, does reasonably well solving 80.5\% of MBPP problems and an even higher 82.3\% of HumanEval tasks, as shown in Figure~\ref{fig:gpt4wofunction}, both of which consist of function-level benchmarks. Section~\ref{file-level} discusses file-level generation, which is inherently more challenging, and also how using another LLM affects these results. These numbers are slightly lower than have been reported for GPT-4 Turbo v1106 in popular LLM coding benchmark leaderboards \cite{liu2024your} by about 5\%. The drop in performance might be attributed to the removal of sample test cases if present in the problem statement for this experiment and differences in the prompt and model parameters like seed and temperature.

\subsection{\RqTwoText{}}
For MBPP and HumanEval, including tests contributes to solving an additional 12.0\% and 8.5\% of problems, respectively, showcasing that tests as part of the input help the LLM reach the correct solution. These percentages correspond to an additional 51 problems in the MBPP dataset and 15 additional problems in the HumanEval dataset. Manually inspecting the results of the evaluation with and without tests, several key insights emerged:

\begin{itemize}
\item \textbf{Problems where the initial implementation had correct logic but incorrect function signatures}: benefited from tests as they helped identify mismatches and ensure adherence to the required function signatures (e.g., MBPP/6, MBPP/90, MBPP/233).
   
\item \textbf{Problems involving mathematical formulas or geometric calculations} saw improvements as tests helped catch errors in formula application and ensured the correct implementation of mathematical logic (e.g., MBPP/14, MBPP/430).

\item \textbf{Problems requiring precise string manipulations or regular expressions} benefited from tests highlighting edge cases and incorrect pattern matching, leading to more robust solutions (e.g., MBPP/16, HumanEval/48).

\item \textbf{Problems involving complex data structures} such as nested dictionaries or tuples. Tests helped ensure that the functions correctly handled various input formats and edge cases. (e.g., MBPP/106, MBPP/299)

\item \textbf{Edge Case Handling}: Many problems initially failed due to unhandled edge cases. Tests were instrumental in identifying these edge cases, prompting the system to refine its logic to handle all possible scenarios (e.g., MBPP/170, HumanEval/5).

\item \textbf{Problems requiring specific algorithmic approaches}, such as finding the maximum difference or sorting criteria, benefited from tests that guided the system to correct logical flaws and align with the problem requirements (e.g., MBPP/63, HumanEval/127).
\end{itemize}
 
LLMs seem to be able to improve their understanding of the requirements presented in the problem statement using the provided tests. The tests also seem to have played a significant role in validating mathematical logic, ensuring the correct application of formulas to produce results indicated by the tests. Tests also seem to prompt the system to refine its logic to accommodate a broader range of inputs and edge cases. Consistency in data types, especially in output formats, was another area where tests highlighted discrepancies. In some cases, adding tests led to simplifying logic, replacing initial complex approaches presumably derived from misunderstanding the statement with more straightforward solutions.

\subsection{\RqThreeText{}}
The remediation loop's role as a corrective mechanism can also be observed, with further improvements noted in problem-solving. For MBPP and HumanEval, the loop seems to effectively refine the code, with an additional 2.8\% and 3\% of problems solved, respectively. This corresponds to the correct code for 21 additional problems for MBPP, out of which 11 were remediated in the first attempt, seven on the second, two on the third and one on the fourth. For the HumanEval dataset, nine additional problems were solved correctly, of which seven were remediated on the first attempt and one each was remediated on the second and third attempts. It should be noted that in our evaluation pipeline, we allowed for up to five remediation attempts. Although a few cases utilized all five attempts, none were solved. Manually inspecting the data, we find the following cases that are benefited by remediation:

\begin{itemize}
\item \textbf{Edge Case Handling}: Many problems initially failed due to inadequate handling of edge cases. Remediation loops allowed the system to incorporate specific checks for these cases, leading to correct code generation (e.g., MBPP/92, MBPP/113, MBPP/137)
\item \textbf{Issues related to data type handling and logical errors} were effectively addressed through remediation loops (e.g., MBPP/115, HumanEval/22)
\item \textbf{Problems requiring precise mathematical computations} saw significant improvements through remediation. Initial attempts often failed due to issues like integer vs. floating-point division, but remediation advice helped correct these errors (e.g., MBPP/300, HumanEval/77)
\item \textbf{Problems involving text manipulation}, such as punctuation handling, benefited greatly from remediation loops. Initial attempts often failed due to improper handling of punctuation, but iterative refinements led to progressively better solutions (e.g., MBPP/7)
\item \textbf{Problems requiring multiple remediation attempts} highlight the importance of iterative refinement. Errors with boundary checks, simple arithmetic corrections, or basic input validation were often resolved through iterative remediation. Each attempt built on the previous one, incorporating new insights and corrections until the correct solution was achieved (e.g., MBPP/160, HumanEval/57)
\end{itemize}

The general trajectory from initial to final remediation attempts involved a progressive refinement process. In most cases, the first remediation led to basic fixes and the handling of straightforward errors highlighted by the crashes. This often addressed the most obvious errors, such as missing imports or missing type conversions. In the case of a second remediation attempt, more detailed adjustments were involved, addressing deeper logical issues and additional edge cases. A few examples of cases remediation attempts to fix are adjusting modulo operations to map to the correct character range, correcting base cases and loop logic in recursive functions, and handling edge cases like empty arrays or specific input formats.

The remediation agent did not have access to the history to prevent the context length of the models from being exceeded during the evaluation, which means the LLM could not review past fixes. There were cases where the remediation advice misled the generator and often failed to fully implement the remediation advice, leading to repeated errors. We ignore cases like MBPP/462 and HumanEval/145, which occurred due to verifier and environment-related issues, as these are limitations of the evaluator as implemented presently. An interesting question is what kind of problems remain unsolved, manually inspecting we find the following:

\begin{itemize}
    \item \textbf{Misunderstanding core logic or requirements} often led to persistent errors across multiple remediation attempts (e.g., MBPP/83, HumanEval/38, MBPP/138, HumanEval/83, MBPP/765).

    \item \textbf{Handling specific data structures} like empty arrays, negative numbers, nested lists, and tuples remained challenging despite multiple iterations (e.g., MBPP/119, MBPP/559, MBPP/431, MBPP/630).

    \item \textbf{Logical errors and inadequate adaptation of remediation advice} were common. Despite following remediation advice, the system consistently produced incorrect logic in cases such as rotating binary strings, renaming functions, and counting and pairing differing bits often led to repeated mistakes (e.g., MBPP/109, MBPP/468, MBPP/126, MBPP/595).

    \item \textbf{Misinterpreting input/output formats and complex cases} involving multiple conditions or constraints frequently caused failures. These included handling nested list inputs and incorrectly handling the input format for date validation (e.g., MBPP/278, MBPP/427, MBPP/102, MBPP/581).

    \item \textbf{Performance bottlenecks and inefficiencies} were encountered, particularly with large inputs and complex logic, leading to timeouts and overly complex solutions (e.g., MBPP/765, HumanEval/63).
\end{itemize}

\subsection{\RqFourText{}}

\begin{table*}[htbp]
\centering
\caption{Impact of private tests on GPT-4 based TGen results with improvement offered by tests after validation highlighted}
\begin{tabular}{lcccccc}
\hline
\textbf{Category} & \multicolumn{2}{c}{\textbf{MBPP(\#399)}} & \multicolumn{2}{c}{\textbf{HumanEval(\#164)}} & \multicolumn{2}{c}{\textbf{CodeChef(\#1100)}} \\
\hline
Solved without tests & 69.67\% &  & 78.66\% &  & 23.00\% &   \\
Needs tests & 82.45\% & \textbf{(+12.78\%)} & 87.81\% & \textbf{(+9.15\%)} & 26.09\% & \textbf{(+3.09\%)}  \\
Needs remediation & 87.71\% & \textbf{(+5.26\%)} & 93.30\% & \textbf{(+5.49\%)} & 30.27\% & \textbf{(+4.18\%)}  \\
Unsolved & 12.28\% &  & 6.71\% &  & 69.73\% &   \\
\hline
Improvement using tests && \textbf{18.04\%} && \textbf{14.64\%} && \textbf{7.27\%}  \\
\end{tabular}
\label{tab:private_tests_impact_across_benchmarks}
\end{table*}

An interesting question in our study is whether the generated solutions are tailored merely to pass the provided tests or if they are robust enough to succeed against unseen tests as well. This question becomes particularly pertinent when considering the potential for omitted requirements in the supplied test cases. While TDD advocates for the use of tests as the primary specifications, the risk of overlooking certain criteria due to suboptimal test quality cannot be ignored.

For the MBPP and HumanEval datasets, we employed their EvalPlus \cite{liu2024your} enriched variants with 35x and 80x more tests, respectively, offering a rigorous assessment platform. EvalPlus seeds its test generator with high-quality inputs and extends them through type-aware mutation to establish a robust framework for evaluating the true correctness of LLM-synthesized code.

Our extended evaluation of the generated code utilizing private tests, summarized in the first two columns of Table \ref{tab:private_tests_impact_across_benchmarks}, indicates that including public tests still results in improved performance across benchmarks. Even though there is a reduction in the number of problems solved corresponding to each category once verified, the addition of public tests yielded a performance increase of 12.78\% for MBPP and 9.15\% for HumanEval. Further application of remediation loops led to additional improvements: 5.26\% for MBPP and 5.49\% for HumanEval. This demonstrates that even without the LLM having access to the comprehensive test suites provided by EvalPlus, including some tests, it aids in enhancing the correctness of the generated solutions over not using tests.



The occurrence of failures in some solutions when subjected to private tests highlights a crucial point: the more exhaustive the test suites are, the better the results tend to be. Our findings reveal that solutions developed with test information fare well against private tests, underscoring the benefits of integrating tests in the code generation process to bolster robustness. This effectiveness points to a nuanced balance: while incorporating tests, even partially comprehensive ones, bolsters solution robustness, the ultimate success hinges on the quality of these tests. Having a wide variety of tests ensures solutions are not just tailored to meet the explicit criteria in the problem statement but are also able to capture implicit requirements.

\begin{figure}
    \centering
    \includegraphics[width=1\linewidth]{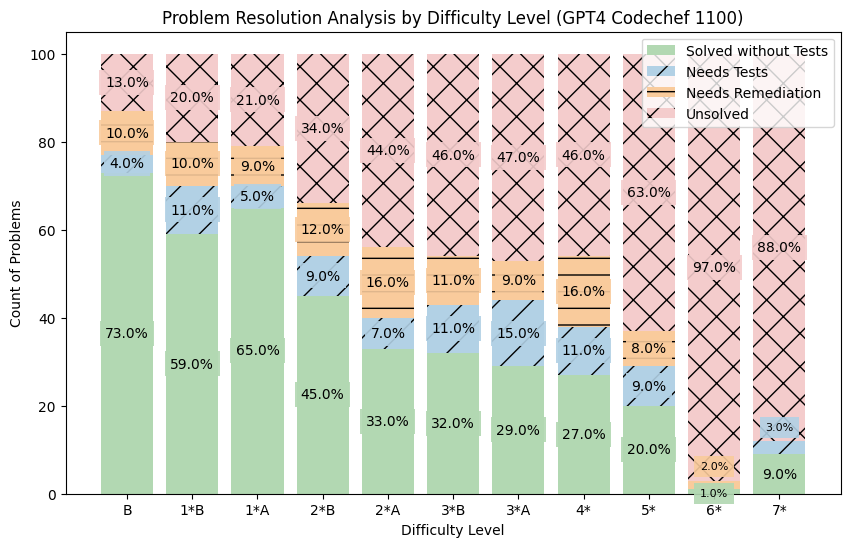}
    \caption{Problems resolved and strategies used compared across difficulty levels using GPT4}
    \label{fig:gpt4wo}
\end{figure}

\section{PARAMETERS AFFECTING TDD}
In this section, we explore the “HOW” question: how do different parameters impact the effectiveness of tests in code generation? We aim to understand the influence of problem difficulty, the number of tests, and the LLM itself. RQs 5 through 7 address these aspects in detail. RQ5 examines how problem difficulty affects code generation results. RQ6 investigates the optimal number of tests needed for effective code generation. RQ7 evaluates whether the results hold when using an open model like Llama 3 as the reasoning engine. This section builds on the findings from Section~\ref{impact}, providing a deeper understanding of the factors that influence the success of integrating TDD with LLM-based code generation.

\subsection{\RqFiveText{}} \label{file-level}

\subsubsection{Data collection}
We curated a problem set of coding problems from CodeChef and their corresponding human solutions. CodeChef assigns each problem on its platform a difficulty score and classifies ranges of difficulties into 11 buckets, as depicted in Table~\ref{tab:problems_diff_level}. To ensure that our study's problem set has a good distribution with respect to difficulty, we fetched the 100 most popular problems from each difficulty level. Here, popularity refers to the number of accepted solution attempts that existed when the data was scraped (November 2023). The curated multi-level dataset of 1100 problems represents a spectrum of real-world coding challenges of varied difficulties that require full program synthesis.

 \begin{table}[h]
\centering
  \caption{Difficulty Levels of Selected CodeChef Problems}
  \label{tab:problems_diff_level}
  \begin{tabular}{rcc}
    \toprule
    Level       &   Range        &  Count   \\
    \midrule
    Beginner    &   0 - 999       & 100      \\
    1* Beginner &   1000 - 1199   & 100     \\
    1* Advanced &   1200 - 1399   & 100     \\
    2* Beginner &   1400 - 1499   & 100     \\
    2* Advanced &   1500 - 1599   & 100     \\
    3* Beginner &   1600 - 1699   & 100     \\
    3* Advanced &   1700 - 1799   & 100     \\
    4*          &   1800 - 1999   & 100     \\
    5*          &   2000 - 2199   & 100      \\
    6*          &   2200 - 2499   & 100     \\
    7*          &   2500 - 5000   &  100     \\
  \bottomrule
                &                 & 1100
\end{tabular}
\end{table}

\subsubsection{Results}

A breakup of the problems and solution strategies is shown in Figure~\ref{fig:gpt4wo}. The data shown can be summarized as 35.72\% (393) problems solved without providing tests, 7.72\% (85) additional problems solved with public tests being supplied to the LLM, and another 9.36\% (103) problems solved using the LLM to look into validation issues and remediating the code. In contrast to function-level generation, where we can expect the function signature and input and output schemes to be straightforward, file-level generation, as required for the curated CodeChef dataset, allows for more flexibility in problem-solving approaches. This flexibility can make CodeChef problems harder to solve, even without considering task complexity. Performance on the CodeChef dataset is markedly lower, with success rates sharply decreasing as problem difficulty escalates. Despite this, 35.72\% of these problems were solved correctly with just the problem statement. This highlights the challenges associated with these problems as compared to function-level generation datasets we have explored so far. We need to parse the input correctly and provide the output to the console in the expected format, apart from the logical complexity of the tasks.

Providing test cases led to a consistent increase in the number of problems solved across difficulty levels, increasing the total number of problems solved by 7.72\%. This underscores the importance of test cases in aiding code generation across all difficulty levels.

Remediation loops further enhance performance in the CodeChef dataset, with an improvement of 9.36\%. This indicates the LLM's capacity for self-improvement and its ability to tackle increasingly sophisticated challenges, particularly for medium-difficulty problems. The complexities associated with correctly handling input and output operations likely contribute to the higher impact of remediation on full program synthesis required for CodeChef tasks.
Most of the CodeChef problems (47.18\% or 519 problems) remained unsolved in our testing configurations. Allowing more remediation attempts could potentially improve these numbers, but our experiments show diminishing returns with increased iterations.

We consider private tests for additional validation to gauge the impact of supplied tests. These tests, not disclosed beforehand, are used by grading systems on platforms like CodeChef to determine correctness. We leveraged these private test suites to evaluate the generated solutions by evaluating the solutions on the platform itself. The results can be seen in the last column of Table~\ref{tab:private_tests_impact_across_benchmarks}. We note that we still observe a 7.27\% improvement in the correctness of solutions across the dataset comprising 1100 problems.


\begin{figure}
    \centering
    \includegraphics[width=1\linewidth]{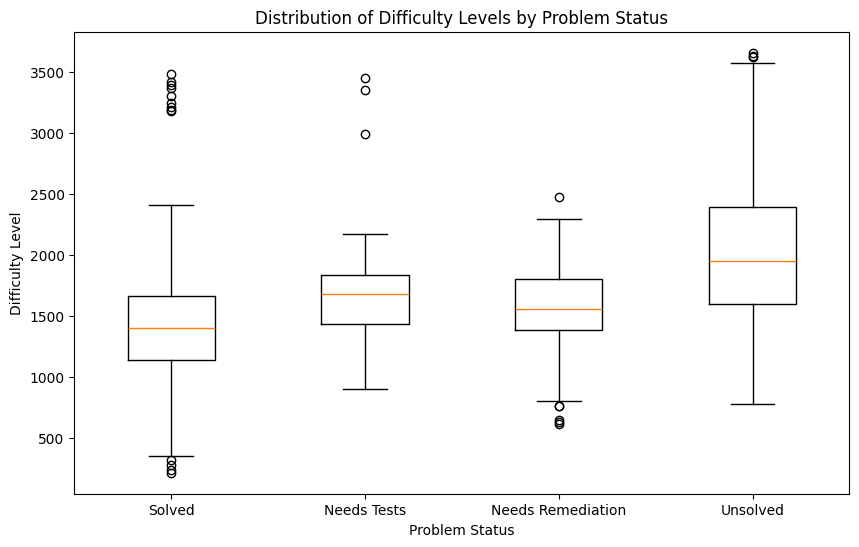}
    \caption{Distribution of difficulty levels associated with problems solved by each strategy}
    \label{distribution}
\end{figure}

Our analysis, as illustrated in Figure~\ref{distribution}, delves into the relationship between problem difficulty and the success rate of generated solutions. The data reveals a clear trend: simpler problems tend to be solved effectively using tests alone, while more complex problems often require additional remediation for successful resolution. Notably, the most challenging problems remained unsolved, indicating a direct correlation between problem difficulty and the likelihood of resolution through our current methodological framework. We see that a much less expensive operation of supplying tests quite significantly impacts the fraction of solved problems. 

Lastly, the discrepancy observed between the success of solutions on public tests versus their failure on CodeChef's private tests raises important concerns. Our analysis of difficulty scores for problems that failed only on private tests yielded an average score of 1627.75, situating these problems within the medium difficulty range. The performance of GPT-4 generated solutions on these tests, where 16 were partially accepted, two caused judge errors, six resulted in runtime errors, and 25 exceeded time limits—underscores the necessity of comprehensive testing. These findings highlight the importance of developing more robust tests that can anticipate and mitigate such issues despite efficiency not being the primary focus of this investigation.

\subsection{\RqSixText{}}

\begin{figure}
    \centering
    \includegraphics[width=1\linewidth]{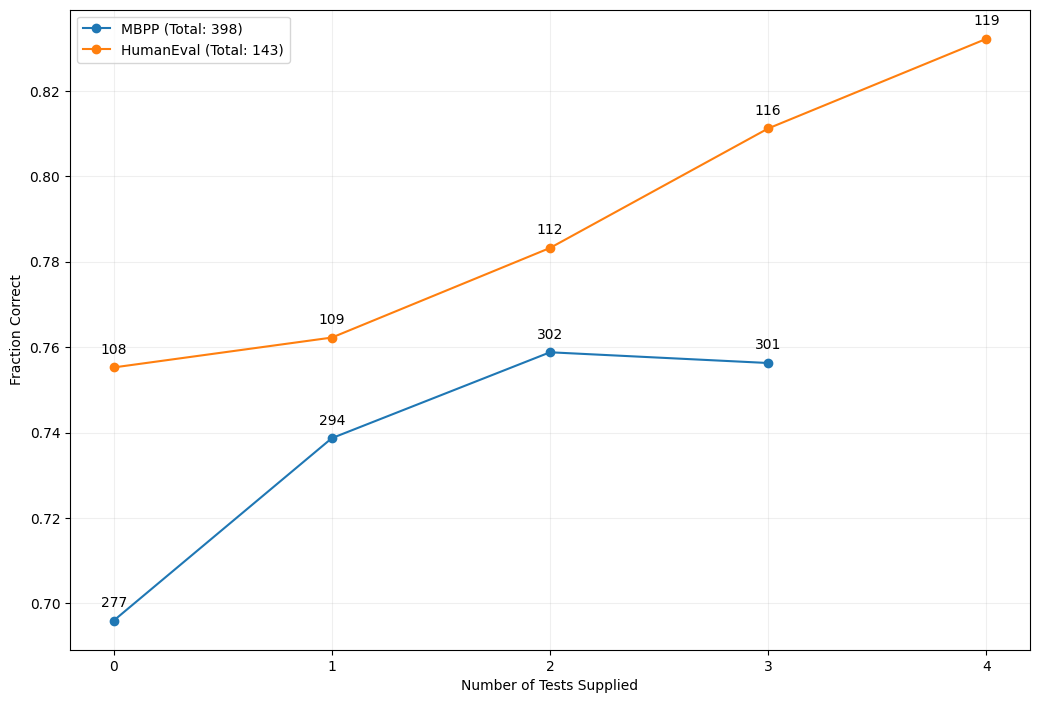}
    \caption{Correctness with increasing number of tests supplied}
    \label{fig:increment_correctness}
\end{figure}

The number of test cases available per problem varies widely across both function-level generation benchmarks. Considering the scope of our study, we choose to include only the human-written tests in these datasets and not those generated by EvalPlus. To explore the impact of the number of these tests supplied to the LLM, we work with a subset of both datasets so that as we vary the number of tests, we are still commenting on the correctness observed for the same set of problems. The cutoffs are picked so that we have enough cases to analyze while not dropping too many problems from our analysis. For MBPP, we look at the subset of problems with at least three tests, resulting in 398 problems, and only one case is removed. For HumanEval, we look for cases with at least 4 tests, leaving us with 143 problems out of the initial 164. 

We run the evaluation pipeline without remediation for each case in these subsets and repeat the experiment while increasing the number of tests supplied. The variation in the fraction of correct solutions generated in each case is shown in Figure~\ref{fig:increment_correctness}. We see that the impact varies across the datasets, but we see an upward trend in both cases as more tests are added. However, we do not observe a plateau in the case of HumanEval, which might be attributed to the simpler nature of problems in MBPP, which seem to suffer more from vagueness than complexity. Also, we note from our experiments that as we continue increasing the number of tests, in some cases, we end up with incorrect code being generated that had been correctly solved before. This could be because of the ``lost in the middle'' problem reported in the literature, where the LLM starts ignoring context in the middle as more and more information is supplied to the model.

Additionally, we look at the sample solution to each of the problems included in the datasets and use it to compute line coverage contributed by each test case. We note that 75\% of the cases in HumanEval and 92.4\% of cases in MBPP achieve 100\% coverage with the addition of the first case itself. This could be another reason why the curve for MBPP is starting to decline with just three tests. We need a better dataset with more human-written test cases for more rigorous analysis, but we hypothesize, based on our results, that adding more diverse test cases aids in the code generation process.

\subsection{\RqSevenText{}}



Apart from GPT-4 Turbo, we also experimented with GPT-3.5 Turbo v1106 and observed similar trends. However, in our discussion so far, we presented only results from GPT-4, as it represents one of the most powerful LLMs available for public use. An interesting question is whether the benefits of TDD can be seen outside OpenAI's GPT family of models. To address this question, we pick Meta Llama 3 70B Instruct, another popular model that shows promise for code generation, ranking among the best models in code \cite{du2024evaluating, liu2024your} and non-code LLM benchmark leaderboards \cite{chiang2024chatbot}. The output from \textit{TGen} for GPT-3.5 Turbo is also included in our replication package.

\begin{figure}
    \centering
    \includegraphics[width=1\linewidth]{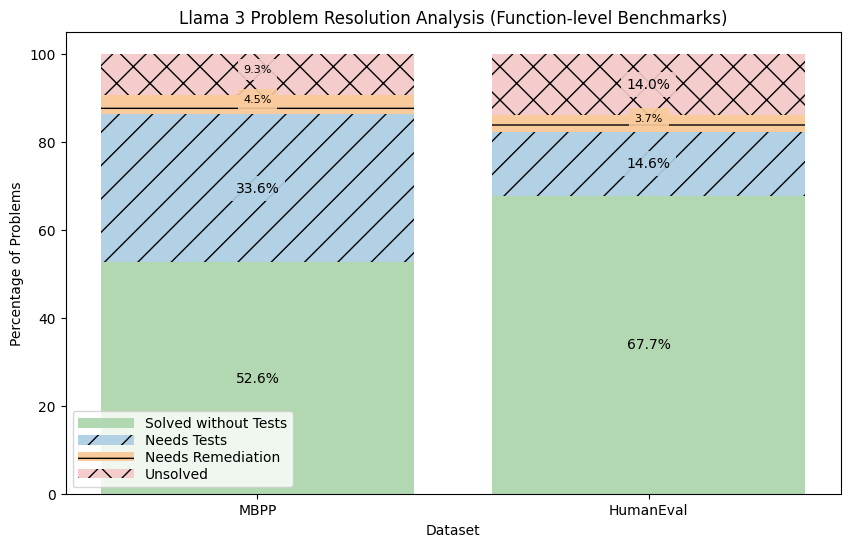}
    \caption{Llama 3: Problems resolved and strategies used for Function-level code generation benchmarks}
    \label{fig:llama3wofunction}
\end{figure}

The results observed by replacing GPT-4 with Llama 3 in the evaluation pipeline can be seen in Figure \ref{fig:llama3wofunction}. 
We see that the trends observed with proprietary models extend to open models. Code generation using Llama 3 benefits from the additional tests and remediation setup provided by \textit{TGen}, and we even see a greater impact than we observed with GPT-4. As compared to GPT-4, which had over 80\% of the problems being solved in the baseline, we see a lower performance of 52.6\% and 67.7\% in MBPP and HumanEval benchmarks, respectively. Despite the worse code generation performance out of the box, we note that the addition of tests and remediation both significantly improve the results. 

The final measure of improvements observed after evaluating the correctness of generated solutions using the EvalPlus can be seen in Table~\ref{tab:llamaprivate_tests_impact_across_benchmarks}. We see that after validating the solutions on private tests, we see a 38.6\% improvement in MBPP and a 21.95\% improvement in HumanEval as compared to their respective baselines, which is almost double the improvement we saw in the case of GPT-4 shown in Table~\ref{tab:private_tests_impact_across_benchmarks}.

\begin{table}[htbp]
\centering
\caption{Impact of private tests on Llama 3 based TGen results with improvement offered by tests after validation highlighted}
\begin{tabular}{lcccc}
\hline
\textbf{Category} & \multicolumn{2}{c}{\textbf{MBPP(\#399)}} & \multicolumn{2}{c}{\textbf{HumanEval(\#164)}}  \\
\hline
Solved without tests & 46.37\% &  & 62.20\% &   \\
Needs tests & 75.94\% & \textbf{(+29.57\%)} & 75.61\% & \textbf{(+13.41\%)} \\
Needs remediation & 84.96\% & \textbf{(+9.02\%)} & 84.15\% & \textbf{(+8.54\%)} \\
Unsolved & 15.04\% &  & 15.85\% &  \\
\hline
Improvement && \textbf{38.60\%} && \textbf{21.95\%}\\
\end{tabular}
\label{tab:llamaprivate_tests_impact_across_benchmarks}
\end{table}

\section{RELATED WORK}
\textbf{Evaluating code generation} performance is an area that is being actively explored. Liu et al.'s EvalPlus \cite{liu2024your} proposes rigorous benchmark enhancements for evaluating LLM-generated code's functional correctness, highlighting the insufficiency of current benchmarks in depth and breadth. Ding et al. present CrossCodeEval \cite{ding2023crosscodeeval}, which challenges LLMs with multilingual, cross-file context code completion. Jimenez et al.'s SWE-bench \cite{jimenez2023swe} uses real-world GitHub issues to test LLMs, with results showing current models solve only a fraction of problems. 

\textbf{Improving code generation using tests} has emerged as a popular trend among other works that have explored other methods, such as improving the reasoning ability of LLMs by using diverse prompts, voting, and other reasoning verification strategies apart from tests \cite{li2023making}. Wang et al. employ test execution feedback while training the model to enable the model itself to discern incorrect code \cite{wang2022test}. Chen et al.'s Codex \cite{chen2021evaluating} demonstrates the limitations of single-sample code generation and suggests that one could similarly exploit multiple test iterations to refine code quality. There have also been attempts at making compilable code by incorporating feedback from compilers \cite{wang2022compilable}. Shinn et al.'s Reflexion \cite{shinn2023reflexion} method allows agents to learn from past mistakes, emphasizing the need for approaches that can iterate over code generation and testing for improved decision-making in software development. Recent agentic systems like Yuntong et al.'s AutoCodeRover have also explored utilizing test cases for spectrum-based fault localization when retrieving context for code generation \cite{zhang2024autocoderover}. However, most of these explore the problem in a program repair setting.

\textbf{Generating tests} is another way tests for code generation are being explored recently \cite{schafer2023empirical}. Lever \cite{ni2023lever} attempts to rank generations based on trained verifiers derived from the prompt. A comprehensive exploration is presented by Codet, where they assess the quality of generated test cases and their impact \cite{chen2022codet}. Lahiri et al. \cite{lahiri2022interactive} explore the use of test cases for user intent formalization, but they rely on developer feedback for mutating and ranking test and code suggestions effectively using generated tests. Fakhoury et al. build upon this by leveraging user feedback through LLM-generated tests to improve the correctness of generated code \cite{fakhoury2024llm}. 

The aforementioned works showcase the potential for using verification and validation techniques to improve the performance of LLMs across different complexity levels and tasks. Further, they also hint at the efficacy of using LLMs for program repair through feedback-based techniques. None of these studies, however, answers key questions surrounding the impact of including human-written test information and how these mechanisms can be employed in a test-driven workflow. We wish to bridge this gap and structure our experiments to explore how to use tests efficiently and how test-driven development principles can be used to improve code generation.

\section{THREATS TO VALIDITY}

Prompt engineering, while a powerful tool to guide LLMs, is also subject to limitations. Poorly designed prompts can lead to sub-optimal results, skewing the model's results and potentially misleading it. To minimize this, we keep the prompt structure consistent across our experiments and vary only the parameters under consideration. Using popular benchmark datasets, such as MBPP and HumanEval, has inherent limitations. These benchmarks do not capture the full complexity and variability of real-world programming problems but help shed light on the impact of tests. We also use our own dataset of problems with a clear notion of difficulty, but this might not be representative of the difficulty of tasks in software engineering. 

Our assumption that test runs serve as ground truths for the corresponding problems may not always hold, particularly for non-deterministic problems. We excluded such problems from our curated dataset and filtered out errors through manual inspection. Recent literature also suggests that the tests used in widely adopted benchmarks may not be robust enough for evaluating code generation. To address this, we used the refined versions of the HumanEval and MBPP datasets curated by EvalPlus \cite{liu2024your}, which include additional tests for more thorough evaluation. The performance of TDD with LLMs might vary with different models, architectures, or training data. Further, the inherent variability in LLM outputs can introduce inconsistencies. We employ a fixed seed value and temperature setting to improve the reproducibility of our results, but some degree of unpredictability remains.

\section{CONCLUSION}

In the rapidly evolving landscape of LLMs for code, where these models are being increasingly used for code generation and remediation, embracing TDD emerges as a strategic paradigm shift. TDD enables us to use tests in the generation phase and for validation. By incorporating test cases and employing remediation loops, we are able to solve complex problems that the LLM cannot solve normally. Using GPT-4, we observe a significant increase in the correctness of the code generated. We find that this improvement is more pronounced for less performant models that can only solve a much lower fraction of problems initially. We also highlight the cases in which these approaches work well and the kind of tasks that present-day LLMs struggle with. Our experiments with GPT-4 and Llama 3 show that merely providing the LLM with tests during generation improves correctness by 9.15 to 29.57\%. Adding remediation loops results in an additional gain of 5.26 to 9.02\% across popular function-level benchmark datasets after rigorous evaluation with private tests. We also find that the proposed approach improves correctness by 7.27\% on our dataset of 1100 file-level generation tasks where GPT-4 performs poorly and struggles with more difficult problems. Our findings validate the impact of tests, and we therefore advocate for the widespread adoption of test-driven methodologies to maximize the benefits of LLMs in code generation.


\bibliographystyle{ACM-Reference-Format}
\bibliography{sample-base}

\end{document}